\documentclass[12pt,a4paper, nofootinbib]{revtex4-1}
\pdfoutput=1
\usepackage[latin1]{inputenc}
\usepackage{amsmath}
\usepackage{footnote}
\usepackage{amsfonts}
\usepackage{amssymb}
\usepackage{graphicx}
\usepackage{graphicx,color}
\usepackage{dcolumn}
\usepackage{multirow}
\usepackage{rotating}
\usepackage{verbatim}
\usepackage{bm}
\usepackage{color}
\usepackage{soul}
\usepackage{float}
\usepackage{comment}
\begin{document}

\title{Studying low-$x$ structure function models with astrophysical tau neutrinos: double bang, lollipop and sugardaddy topologies}

\author{D. A. Fagundes}
\email{daniel.fagundes@ufsc.br}
\affiliation{Departamento de Ciências Exatas e Educação, Universidade Federal de Santa Catarina - Campus Blumenau, CEP 89065-300, Blumenau, SC, Brazil}

\author{R. R. Francisco} 
\email{rafael.francisco@udesc.br}
\affiliation{Departamento de Física, Universidade do Estado de Santa Catarina - Campus Joinville, 89219-970, Joinville, SC, Brasil. \\
Departamento de Engenharia do Petróleo, Universidade do Estado de Santa Catarina - Campus Balneário Camboriú, 88336-275, Balneário Camboriú, SC, Brasil.}

\author{E. G. de Oliveira} 
\email{emmanuel.de.oliveira@ufsc.br}
\affiliation{Departamento de F\'{i}sica, CFM, Universidade Federal de Santa Catarina, C.P. 476, CEP 88.040-900, Florian\'opolis, SC, Brazil}

\date{\today}

\begin{abstract}
Despite not have been yet identified by the IceCube detector, events generated from $\nu_{\tau}$ deep inelastic neutrino scattering in ice with varied topologies, such as double cascades (often called \textit{double bangs}), \textit{lollipops} and \textit{sugardaddies}, constitute a potential laboratory for low-x parton studies. Here we investigate these events, analyzing the effect of next-to-next-to-leading order (NNLO) Parton Distribution Function (PDFs) in the total neutrino--nucleon cross section, as compared with the color dipole formalism, where saturation effects play a major role. Energy deposit profiles in the `bangs' are also analysed in terms of virtual $W$-boson and tauon energy distributions and are found to be crucial in establishing a clear signal for gluon distribution determination at very small $x$. By taking the average (all flavor) neutrino flux ($\Phi_{\nu}\sim E_{\nu}^{-2.3}$) into differential cross sections as a function of  $\tau$ and $W$ energies, we find significant deviations from pure DGLAP parton interactions for neutrino energies already at a few PeV. With these findings one aims at providing not only possible observables to be measured in large volume neutrino detectors in the near future, but also theoretical ways of unravelling QCD dynamics using unintegrated neutrino-nucleon cross sections in the ultrahigh-energy frontier.

\end{abstract}
\maketitle

\section{Introduction}
Neutrino physics has been achieving significant results and development in the last twenty years~\cite{Ackermann:2019cxh}, such as the first detection of tau neutrinos~\cite{Kodama}, the confirmation of neutrino oscillations~\cite{YFukuda}, etc. Big part of these achievements were attainable just after the construction and improvement of several neutrino detectors as the Super-Kamiokande~\cite{SFukuda}, IceCube~\cite{Aartsen:2017sml}, MinibooNE~\cite{Green}, among others that allowed more detailed studies of neutrino interactions at different energy scales. Currently, there are many proposals {for} new neutrino detectors or upgrades in the ongoing experiments like IceCube-Gen2 (South Pole)~\cite{Aartsen:2014ice2}, GVD (Baikal Lake)~\cite{AD-Avrorin} and KM3NeT (Mediterranean Sea)~\cite{Spiering:2017kkh}. Some of them are already in construction or have recently started operating, such as KM3NET. 
This new generation of neutrino telescopes shall improve statistics and sensitivity by roughly one order of magnitude. 

IceCube, located in Antarctica, is a large volume of ice (more than $1$\,km$^{3}$) that acts as a Cherenkov-light detector. Its main purpose is to detect high-energy neutrinos when they scatter off the ice. Since its construction was finished it has already observed dozens of high-energy neutrinos with energies above $100$\,TeV. Therefore, IceCube and similar detectors offer a wonderful possibility of studying high energy neutrino--nucleon collisions and, as a result, to have a better understanding of the proton structure. Most of the high energy neutrinos will be of astrophysical origin and comprise electron, muon and tauon neutrinos, with expected flavor ratio of 1:1:1~\cite{Aartsen:2015ivb}, since the neutrino oscillations average the ratio at the source in most accepted scenarios.

Neutrinos can interact with the ice through the exchange of a Z boson. In these neutral current interactions, the neutrino is in the final state and carries part of its initial energy. The energy that is deposited in the ice (i.e.\ boson energy) will produce relativistic charged particles that in turn can produce Cherenkov radiation, which is actually what IceCube can detect. These neutral current events will produce a signature (or topology) called ``shower'', since they are a shower of produced particles. In these events, the neutrino flavor plays little to no role. 

Much more common is the charged current (CC) interaction, in which a W boson is exchanged. In this case, the neutrino becomes an electron (or positron), muon or tauon after the interaction. Therefore, there is Cherenkov light emitted by the generated lepton and the secondary relativistic charged particles. When electron or positron is created, they do not travel a long path inside the ice, since they readily interact with the medium. As such, their signature is also a shower of particles. However, muons travel longer in detectors than electrons, since they can more easily penetrate material. In doing so, a considerable amount of Cherenkov photons is emitted as they travel, and this signature is called appropriately a ``track''.  Therefore, the detector can fairly distinguish muon neutrinos that interact by charged current.

The remaining possibility is the production of a tauon, that can also penetrate the medium like the muon. However, it decays much faster than muons and, if it has low energy, what is usually detected is the combination of tauon decay with the $W$ boson interaction with the ice on the same spot, making it very hard to distinguish from a regular electron or positron shower. However, if the tauon energy is of the order of hundred TeV or larger, it usually will go through a significant distance from the position where the neutrino interacted with matter until the decay position. It is estimated~\cite{Aartsen:2015dlt} that the average length of tauon decay scales with energy as roughly $5\,$cm/TeV.

\begin{figure}[htb]
\begin{center}
 \includegraphics[width=9cm,height=5.cm]{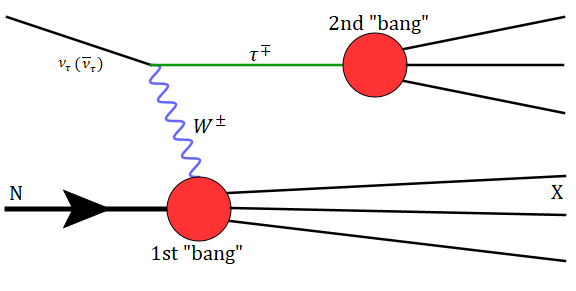}
\caption{\sf\small Double-cascade event diagram representing, at the left upper vertex, how a very high-energy tau (anti)neutrino becomes a (anti)tauon through the emission of its respective charged vector boson. The first red `blob', at the left lower vertex, corresponds to the first ``bang" generated once $W^{\pm}$ interacts with a nucleon. Likewise, at the right upper vertex, the second red `blob' specifies  the second ``bang'' occurring right after the $\tau$ lepton decays.}
\label{fig:db}
\end{center}
\end{figure}

Consequently, a high energy tauon generates a signature different than the regular ``shower" or ``track''. In a Cherenkov detector like IceCube, this pattern can be identified like two separated showers generated from the tau neutrino interaction with the ice and the tauon decay, linked by a muon-like Cherenkov trace, corresponding to the path the tauon went through, as shown in Fig.~\ref{fig:db}. This event is commonly referred to in the literature as a ``double-bang" event \cite{Aartsen:2017sml}. 

In the case one of the bangs happens outside the detector, a different Cherenkov-photons topology has to be identified. When the first bang happens outside the detector, the event is called a ``lollipop" if the tauon decays to hadrons where we can just observe a track and a bang. If the tauon decays into muon, the event is identified as a ``sugardaddy" topology. This one can be observed as a track crossing all the detector length that starts emitting few photons (``thin'') and then suddenly becomes wider after the tauon decay. When the second bang happens outside the detector, it is a ``inverted lollipop" topology. The biggest problem in detecting inverted lollipops is to distinguish them from regular muon tracks. The main topologies are reviewed in Ref.~\cite{Cowen:2007ny}. 

Since the production of tauons provides more information about the neutrino--matter interaction than the more simpler shower or track events, we will consider the following process in this paper:
\begin{equation}
\label{eq1}
\nu_{\tau}+N\rightarrow \tau+X.
\end{equation} 
Our main goal is to show how observables built from this reaction can provide information about nucleon structure, and we do this by using current models for the process as examples. Although these events were not identified on IceCube yet, it is expected that they should be detected soon, as  the current probability to detection is of 90\% at IceCube~\cite{Palladino:2018qgi}. With the further developments on new detectors this could become much more common in next years \cite{Aartsen:2014ice2, Giaz, Brailsford, Capozzi}. 

An approach we investigate is to determine the differential neutrino cross section on tauon (or $W$ boson) energy, if it is possible to detect both bangs. Following this reasoning, the differential cross section is calculated from two different models, one using the parton distribution functions (PDFs) in a extended perturbative analysis and another from the color dipole model. The same idea can be used for the topologies in which only one bang is detected, if the astrophysical neutrino flux is known. As far as we know, this last possibility has not been investigated so far in the literature.

The importance of detecting tau neutrinos is to confirm the current picture of neutrino oscillations~\cite{Palladino:2018qgi} and the expected ratio of 1:1:1. Also, the study of third generation particle interactions is frequently used to track signs of new physics (NP)~\cite{Coloma:2017ppo}, and recent experiments performed at the MinibooNE~\cite{Arevalo} and at the LHCb \cite{LHCb, Aaij:2015yra} present results that can suggest indications of NP in process involving tau neutrinos and tauons. A very recent example is the ANITA anomalous events~\cite{Fox:2018syq} that are cosmic ray extensive showers with energies around $10^9\,$GeV, which could in principle be explained as tau neutrino interactions. They are called anomalous since they are observed with a rather large exit angle as measured from the horizon, but tau neutrinos lose energy when travelling through matter and therefore such occurrences should be very rare. The observation of the two events differ from Standard Model predictions at $5.8\, \sigma$ confidence level, suggesting that in order to explain the phenomena, there can be a NP particle that can travel through the Earth to produce such events. 

Starting in the near futrue, IceCube-gen2 is predicted to get around ten times more statistics than IceCube~\cite{Aartsen:2014ice2} and, as such, it will provide some information in the uncharted territory of very small $x$ PDFs. Recent works \cite{Aartsen:2016ngq, Aartsen:2018vtx} have supported the hypothesis of astrophysical origin for IceCube detections above $100$ TeV and future detectors can be built in a way to more conveniently identify tau neutrinos. A few detections ($\sim 5$) of these events could already improve PDFs determination and probe NP, for which our predictions will be useful. 

This paper is organized as follows: in Sec.\ II, after looking at the basics of neutrino--nucleon deep inelastic scattering, we review the PDF model and then the color dipole method, using two proposed dipole cross sections: GBW ~\cite{GBW:1998} and Arguelles et al.\ (AHWKR) ~\cite{Arguelles:2015wba}. In Sec.\ III we present our results; i.e.\  calculate the total and the differential (on W boson energy) neutrino cross sections, comparing the results from the two methods and its implementations. Then, we integrate the differential cross section averaged by the astrophysical neutrino flux. In Sec.\ IV we give a summary and discuss the perspectives of the double-bang signature detection considering the results presented.   

\section{Neutrino--nucleon charged-current deep inelastic scattering}

In this section we review the basics of neutrino--nucleon charged-current deep inelastic scattering that are needed in order to obtain our results, as well as the two models employed. We take the nucleon to be an average of proton and neutron, the so called isoscalar target. We use the proton (nucleon) mass $m_p = 0.938$\,GeV$/c^2$ while the $W$ mass is $m_W = 80.4$\,GeV$/c^2$. We also discuss the neutrino flux at the end of the section.

Let us start with the kinematics. The tau neutrino, the tauon and the nucleon have four-momentum $k$, $k'$ and $P$, respectively. In the charged current interactions, the neutrino interacts with the nucleon through a $W$ boson that carries four-momentum $q = k - k'$ with virtuality $Q^2 = - q^2$. The c.o.m.\ energy squared is given by $s = (k + P)^2$. The other two invariants we choose to work with are the inelasticity parameter $y = q \cdot P / k \cdot P$ and the Bjorken variable $x = Q^2 / (2 P \cdot q)$. The virtuality in terms of these invariants is given by: $Q^2 = x y (s - m_p^2)$.

It is compelling to work in the reference frame where the target is at rest, since this is the lab frame for neutrino detectors. In this frame, the neutrino arrives with energy $E_\nu$, while the nucleon is at rest; the tauon has energy $E_\tau$ and the energy of the $W$ boson is $E_W = E_\nu - E_\tau$. The inelasticity parameter is $y = (E_\nu - E_\tau) / E_\nu$ and, in a double bang event, it is roughly the energy fraction of the first bang, while $1 - y = E_\tau / E_\nu$ is the energy fraction of the second bang. Also in the nucleon rest frame $s = 2 E_\nu m_p + m_p^2$. 

\subsection{Parton model}

The differential cross section for a neutrino--nucleon charged current interaction is given by ~\cite{CooperSarkar:2011pa}:
\begin{align}
\frac{d \sigma}{d y \, dx}
 = 
 \frac{G_F^2 E_\nu m_p}{2 \pi}  \left( \frac{M_W^2}{M_W^2 + Q^2} \right)^2
  (Y_+ F_T 
  + 2 (1-y) F_L
  \pm Y_- x F_3)
  \label{eq:diffXY}
\end{align}
where $Y_\pm  = 1 \pm (1-y)^2$, $G_F = 1.17 \cdot 10^{-5}$\,GeV$^{-2}$, and $F_T$, $F_L$, and $F_3$ are structure functions that depend on $x$ and $Q^2$. The plus (minus) sign is chosen for a incoming (anti)neutrino. For the total cross section, we have to integrate $0 < y < 1$ and $ Q_\text{min}^2/y(s - m_p^2) < x < 1$, where $Q_\text{min}^2 = 1\,\text{GeV}^2$ makes sure perturbation theory is valid. For the differential in $E_W$ cross section, we use $d y = d E_W/E_\nu$. 

In the QCD improved parton model, the nucleon structure is due to its composition in terms of partons. Specifically, for the case of a neutrino scattering off a isoscalar nucleon, at leading order ($F_L = 0$; $F_T = F_2$)~\cite{CooperSarkar:2011pa}: 
\begin{align}
F_T(x,Q^2)  = x (d + u + \overline{u} + \overline{d} + 2 s + 2 \overline{c} + 2 b)
\end{align}
and
\begin{align}
x F_3(x,Q^2) = x (d + u - \overline{u} - \overline{d} + 2 s - 2 \overline{c} + 2 b)
\end{align}
where $d=d(x,Q^2)$ is the down quark parton distribution function, $u$ is the up quark, and so on. 

We will calculate the coefficient functions up to next-to-leading order (NLO), despite that, at high energies, corrections proportional to $\alpha_s$ are very small. However, for the parton evolution, we choose the PDFs at NNLO in order to include $\alpha_s^i \ln^j(Q^2/Q_0^2)$ terms. As a result of this, the $Q^2$ dependence of the parton distributions are accounted in the best way available. The above parton distributions will be taken from the global fit parametrizations CT14~\cite{Dulat:2015mca}, MMHT14~\cite{Harland-Lang:2014zoa} and NNPDF3.1~\cite{Ball:2017nwa} through the LHAPDF~\cite{Buckley:2014ana} package. These parametrizations use the general mass variable flavor number scheme (GM-VFNS). We will also plot the 68\% confidence interval uncertainty band provided by MMHT14, defined using the conventional Hessian approach.

Another issue is the momentum fraction $x$ dependence. As the parametrizations are determined from available data, they are reliably known  for $x \gtrsim 10^{-4}$, e.g., Fig.~2.1 of Ref.~\cite{Ball:2017nwa}. Below this limit, the PDF fitting groups make reasonable extrapolations that may differ. If a very small $x$ is reached, such as there is not an extrapolation provided by the group, the LHAPDF will provide one based on the smallest values of $x$ available. This is an important point since, in this paper, the high energy neutrino will probe a region of very small $x$, for instance, the lower limit in $x$ taking $Q_\text{min} \approx m_p$ will be $x \approx m_p/2 E_W$, and with a PeV neutrino, $x \approx 10^{-6}$ will be probed. In this region, it is very well possible that the 
pQCD DGLAP evolution has to be supplemented by BFKL large $\log(1/x)$ resummation or some other absorptive corrections.

\subsection{Color dipole model}
\label{sec:dipoles}

To include very low-x corrections to the parton model in the present work, we also use the Color Dipole Model (CDM), following up closely two previous studies \cite{Kutak:2003bd,Arguelles:2015wba}. In this approach, the first step is to define the probability density for a virtual boson $W$ to fluctuate into a dipole, i.e., a $\overline{q}q$ pair, considering both transverse and longitudinal polarizations.This wave function is given by the following formulae in massless quark limit (see e.g.~\cite{Kutak:2003bd}):
\begin{eqnarray}
\rho _{T}(z,r,Q^{2})\equiv | \psi_{T}^{W}(z,r,Q^{2})|^{2} &=&\frac{6}{\pi^{2}}(z^{2}+ \overline{z}^{2})\overline{Q}^{2}K^{2}_{1}(\overline{Q} r);\\
\rho _{L}(z,r,Q^{2})\equiv | \psi_{L}^{W}(z,r,Q^{2})|^{2} &=&\frac{24}{\pi^{2}}(z\overline{z})\overline{Q}^{2}K^{2}_{0}(\overline{Q}r);
\end{eqnarray}
where $z (\overline{z}=1-z)$ specifies the fraction of longitudinal momentum of the quark (anti-quark) in the pair and $r$ defines the transverse distance of the dipole. In addition, $K_{0,1}$ are the zeroth and first-order modified Bessel functions and $\overline{Q}^2=z\overline{z}Q^{2}$. 

In this framework, structure functions are evaluated through the following integral \cite{GBW:1998} (with the sum over the number of masless quark flavors implicit):
\begin{equation}
F _{T/L}(x,Q^{2})=\frac{Q^{2}}{4\pi^{2}}\int d^{2}\textbf{r}\int_{0}^{1} dz  \rho _{T/L}(z,r,Q^{2}) \sigma_{d}(x,r)
\end{equation}
where the dipole cross section, $\sigma_{d}(x,r)$, represents the imaginary part of the scattering amplitude of a quark-antiquark dipole, $\overline{q}q$, off a nucleon target. These structure functions are then used in Eq.~\ref{eq:diffXY} with $F_3=0$.

While the expressions for the W/Z wave functions are consensus in the literature, there are many  $\sigma_{d}(x,r)$ parametrizations. For that reason, here we analyze two dipole models, the first one being the well-known GBW dipole model \cite{GBW:1998}, for which:
\begin{equation}
\sigma_{d}^\text{GBW}(x,r)=\sigma_{0}(1-e^{-r^{2}/4R^{2}_{s}(x)}),
\end{equation}
with $R^{2}_{s}(x)=1/Q_{s}^{2}$ as the typical transverse radius of the dipole related to the saturation scale: $Q_{s}^{2}(x)=Q_{0}^{2}(x_{0} / x)^{\lambda}$. The best fit parameters describing $F_{2}(x,Q^{2})$ data from HERA  in the range $x\leqslant 0.01$, with four active flavors are nucleon size $\sigma_{0}=29.12$ mb, $Q_{0}^{2}=1.0$ GeV$^{2}$, $x_{0}=4.1\times 10^{-5}$, and $\lambda=0.277$. 

Despite the fact that heavy quarks can play a role at very high energy, such contribution was found to be small in \cite{Kwiecinski:1998yf}. Therefore, for consistency, we shall assume $n_{f}=4$ throughout our calculations with the GBW model, assuming (as previously mentioned) massless quarks. Moreover, to improve the model and account for the large $x$ region ($x\sim 1$), we correct the structure functions, $F_{T/L}(x,Q^{2})$ by a factor $(1-x)^{2n_{s}-1}$, with $n_{s}=4$ representing the number of sea quark flavors, following from the constituent quark counting rules as suggested in Ref. \cite{Kutak:2003bd}. {This correction provides a screening effect in the dipole cross section for $x\gtrsim 10^{-2}$, where GBW is not supposed to give reliable results\footnote{Specifically, for small $r$ and  $x\sim 0.1$, the dipole cross section decreases by a factor one-half.}.}

The second model investigated is the hybrid pQCD-dipole model {proposed} by Argüeles et al.~\cite{Arguelles:2015wba} (from hereon called AHWKR). Now the standard parton model is applied to compute structure functions $F_{T}$ and $F_{L}$ in the range $1 > x > x_{0}$. For simplicity, we repeat\footnote{However, being aware that results may not be so sensitive to different choices in the range $x_{0}=10^{-2}-10^{-6}$.} the original choice of Ref.~\cite{Arguelles:2015wba}, the arbitrary value $x_{0}=10^{-5}$. Below the cutoff, where absorptive corrections become important, the dipole formalism is invoked with a specific dipole cross section.

In contrast to GBW model, the AHWKR dipole cross section monotonically increase in the small $x$ region (and therefore does not saturate {at large $r$}). In effect, their dipole cross section derives from an approximation for large $Q^{2}$ behaviour of $F_{2}^{\gamma p}(x,Q^{2})$ \cite{Ewerz:2011ph} (see e.g.\ Eq.~4 of Ref. \cite{Arguelles:2015wba}), being roughly given by the logarithmic slope in $Q^{2}$ of $F_{2}^{\gamma p}$
\begin{equation}
\sigma_{d}(x,r) \sim r^{2}\frac{\partial }{\partial \ln Q^{2}} F_{2}^{\gamma p}(x,Q^{2})\left|_{Q^{2}=(z_{0}/r)^2}\right. ,
\end{equation}
with $z_{0}=2.4$. For high virtualities, $F_{2}^{\gamma p}(x,Q^{2})$ is parametrized following the 
Block-Durand-Ha \cite{Block:2014kza} (BDH) procedure, namely,  a Froissart-bounded expression that ensures all hadron cross sections to asymptotically rise as $\sigma \sim \ln^{2}(s/s_{0})$, with $s_{0}$ being an arbitrary high energy scale. With unitarity guaranteed, by imposing a constraint to the asymptotic energy behaviour of $F_{2}^{\gamma p}(x,Q^{2})\sim \ln^{2}(1/x)$, the necessary condition to obtain the aforementioned Froissart bevahiour is achieved. In effect, the BDH approach embodies NLO QCD corrections in $F_{2}$ below the top quark threshold, i.e. for $m_{b}^{2}<Q^{2}<m_{t}^{2}$, thus keeping $n_{f}=5$ \cite{Block:2013nia}. Moreover, as a by-product, charged and neutral current neutrino-nucleon cross sections, are also asymptotically bounded, since at asymptotic energies $\sigma_{\nu N} \sim \ln^{3} E_{\nu}$ (where $E_{\nu}$ is the laboratory neutrino energy). 

Hence, by using the most recent BDH parametrization \cite{Block:2014kza} of $F_{2}^{\gamma p}(x,Q^{2})$, the following cross section is obtained \cite{Arguelles:2015wba}  
\begin{equation}
\label{eq:ahwkr_dip}
\sigma_{d}^{AHWKR}(x,r) = \mathcal{N}\alpha(x,r)\left[\gamma_{0}+\sum_{i=1}^{4}\gamma_{i}(x,r)\right],
\end{equation}
where 
\begin{eqnarray}
\alpha(x,r)&=& \frac{\pi^{3}r^{2}(1-x)^{n}}{(\mu r)^{2}+z_{0}^{2}};\end{eqnarray}
and the $\gamma$ terms are given by:
\begin{eqnarray}
\gamma_{0}&=& c_{1}z_{0}^{2};\\
\gamma_{1}(x,r)&=& a_{0}(\mu r)^{2};\\
\gamma_{2}(x,r)&=& \mathcal{A}(a_{1}z_{0}^{2}+2\mathcal{B}(a_{2}z_{0}^{2}+b_{1}(\mu r)^{2}+\mathcal{B}b_{2}(\mu r)^{2})+2b_{0}(\mu r)^{2});\\
\gamma_{3}(x,r)&=& \mathcal{B}(\mu r)^{2}(a_{1}+a_{2}\mathcal{B});\\
\gamma_{4}(x,r)&=&z_{0}^{2}\mathcal{A}^{2}(b_{1}+2b_{2}\mathcal{B});
\end{eqnarray}
with $\mathcal{A}=\ln\left(\frac{z_{0}^{2}/x}{(\mu r)^{2}+z_{0}^{2}} \right)$ and $\mathcal{B}=\ln\left(1+\left( \frac{z_{0}}{\mu r}\right)^{2}  \right) $. In Eq.~\ref{eq:ahwkr_dip}, $\mathcal{N}=0.71$ is a normalization constant, introduced to match the $F_{2}^{\gamma p}(x,Q^{2})$ fit by BDH with the photoproduction cross section of $\overline{q}q$ pairs, $\sigma_{\gamma* p\rightarrow \overline{q}q+X}$ \cite{Block:2014kza}. This procedure is in order to make a safe interpolation between the validity domain of PDFs ($x>x_{0}$) and dipole models ($x<x_{0}$). The fit parameters of the model, $a_{0},a_{1}, a_{2}$, $b_{0},b_{1},b_{2}$, $c_{1}$, $\mu^{2}$, $z_{0}$ and $n$ follow from Table I of Ref. \cite{Arguelles:2015wba}, except for $n$ (which is given in Table I  of Ref. \cite{Block:2014kza}).

Both dipole models have been derived from the analysis of electron-proton $(ep)$ DIS by the HERA-ZEUS Collaborations in a wide range of $x$ and $Q^{2}$. Evidently, they were tunned by fitting these data in very distinct kinematic domains and under different hypotheses, but the essential information they convey is related to the strong interaction between the dipole and the target (isoscalar) nucleon. Specifically, both fit the $\sigma_{\gamma^{*}p}$ data, factorizing the QED component in the wave functions from the hadronic one, encompassed in $\sigma_{d}$.  In the electroweak sector, where both charged and neutral current processes can be addressed, the hadronic part remains intact, while the proper wave functions of W/Z fluctuations into $\overline{q}q$ pairs must be provided. The choice of wave functions have followed from previous analysis of ultra-high energy neutrino interactions \cite{Arguelles:2015wba,GBW:1998, Kutak:2003bd}, in which the massless quark limit was taken.

The predictions of models GBW and AHWKR for CC observables such as the total cross section, $\sigma
 _{\nu p}$ and W energy distributions, $d\sigma_{\nu p}/dE_{W}$ are given in Section III. Next, we discuss how the {astrophysical neutrinos flux} and the cross section in the context of parton and dipole models can be used to estimate inelasticity profiles of double bangs in IceCube and next generation detectors.

\subsection{Neutrino flux}

Most of the neutrinos at very high energies will be of astrophysical origin, as observations at IceCube~\cite{Aartsen:2016ngq, Aartsen:2018vtx} have disfavoured the hypothesis that such events can be produced { by interaction} of cosmic rays with the cosmic radiation background (cosmogenic neutrinos) or with the atmosphere (conventional and prompt atmospheric neutrinos).
We will assume that the flux composition is $(\nu_e: \nu_\mu: \nu_\tau) = (1:1:1)$, which is appropriate for PeV neutrinos of astrophysical origin. This is a result of averaging over astronomical distances due to neutrino mixing, even if the original sources produce a biased ratio ({see e.g.\ Ref.~\cite{Fargion:2015lmv}, where astrophysical neutrino flux is  discussed at length}). As the current detectors cannot discern between an incoming neutrino and its respective antiparticle, we should consider the average between the $\nu p$ and the $\overline{\nu}p$ interactions. However, as discussed in \cite{Aartsen:2019mid} for energies above 100 TeV the cross sections of $\nu$ and $\bar{\nu}$ are almost identical, so the average gives the same result of just analysing the $\nu p$ interaction. 

Nonetheless, the flux dependence on neutrino energy is of course important, but not completely understood. Here we shall assume, as given in  Fig.~4 of Ref.~\cite{Aartsen:2014gkd}, a power law spectra $\sim E_{\nu}^{-(2+\delta)}$, namely:
\begin{align} \label{eq:flux} 
E_\nu^2 \frac{d f_\nu}{d E_\nu} = \phi_0 \left( \frac{E_\nu}{E_0}\right)^{-\delta}
\end{align}
with $\phi_0 = 1.5 \cdot 10^{-8}$ GeV / (cm$^2$ s sr), $\delta=0.3$ and $E_0=100$\,TeV. In the above expression $f_\nu$ { represents} the neutrino flux per detector area, observation time and solid angle.

While the angular dependence of the flux is nontrivial and important~\cite{Kistler:2016ask}, here  we are mostly interested in the proton structure, so we will use physical quantities and ratios that are not sensitive to it. Experimentally this could be realized considering only downgoing neutrinos to reduce the uncertainty on the flux due to absorption and regeneration~\cite{Kwiecinski:1998yf, Vincent:2017svp}.

\section{Results and Discussion}

\begin{figure} [htb]
	\begin{center}
		\includegraphics[clip=true,trim=0.0cm .0cm 0.0cm 0.0cm,width=12.0cm]{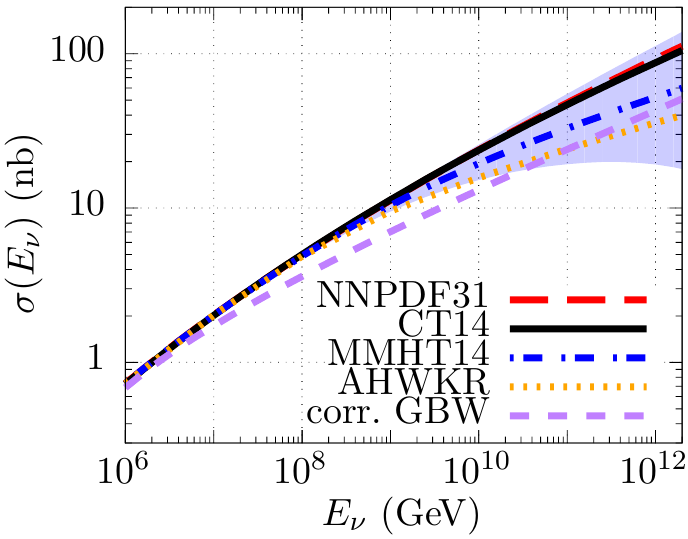}
		\caption{\sf\small Charged-current (CC)   neutrino--isoscalar nucleon ($\nu$--$(p + n)/2$) cross section  as a function of laboratory neutrino energy, $E_{\nu}$, calculated for CT14~\cite{Dulat:2015mca}, MMHT14~\cite{Harland-Lang:2014zoa} (with uncertainty band), and NNPDF3.1~\cite{Ball:2017nwa} NNLO parton distribution functions (PDFs); as well as for AHWKR~\cite{Arguelles:2015wba} and corrected GBW~\cite{GBW:1998} color dipole models. For the hybrid model AHWKR, the perturbative QCD component is switch off for $x<10^{-5}$. }
		\label{fig:nuP}
	\end{center}
\end{figure}

Our results for the total cross section appear in Fig.~\ref{fig:nuP}, where we present the parton model calculations using three distinct NNLO PDFs, namely CT14, MMHT14 (with uncertainty band), and NNPDF31, as well as two color dipole models, AHWKR and corrected GBW. We stress that this Fig.\ is practically the same if we are talking about electron, muon or tauon neutrinos. Up to $E_\nu = 10^9$ GeV, there is a good agreement of the three parton distribution results; in special for CT14 and NNPDF31, for which the curves almost overlap in a wider energy range. At the higher energies, where the low-x behaviour becomes more significant, the difference among the central results can be of about a factor 2, as one notice that MMHT14 extrapolation to $x\lesssim 10^{-6}$ yields a significant smaller cross section. However, it is possible to see that most of the predictions fall within the MMHT14 uncertainty band.

Adding the dipole model into the analysis, one can see significant discrepancies at $E_\nu = 10^9$ GeV already, while at high energies the dipole model results are around 40\% of the largest PDF ones. AHWKR shows a large suppression at $10^{12}$ GeV, while agreeing more with the PDFs at lower energies; for instance, at energies of $E_\nu = 10^8$ GeV or smaller, we see that all curves agree, except the corrected GBW one. We stress that not correcting GBW by a factor $(1-x)^{2n_{s}-1}$ (as explained in Section \ref{sec:dipoles}) might generate an even larger discrepancy already in the energy range of $10^6-10^7$ GeV. 

With our focus in $x$ and $Q^2$ regions where the structure of the proton is not determined from available experiments, one seeks less inclusive observables that can be sensitive to modeling of structure functions in these regions. Here is where the double bangs events can be useful, as both bang energies can be in principle measured. The first bang will be roughly proportional to the $W$ boson energy, while the second bang will be related to the tauon energy. As we shall show, variations in this observable may be truly important already at lower energies, since it is a less inclusive result. 

\begin{figure} [htb]
	\begin{center}
		\includegraphics[clip=true,trim=0.0cm .0cm 0.0cm 0.0cm,width=12.0cm]{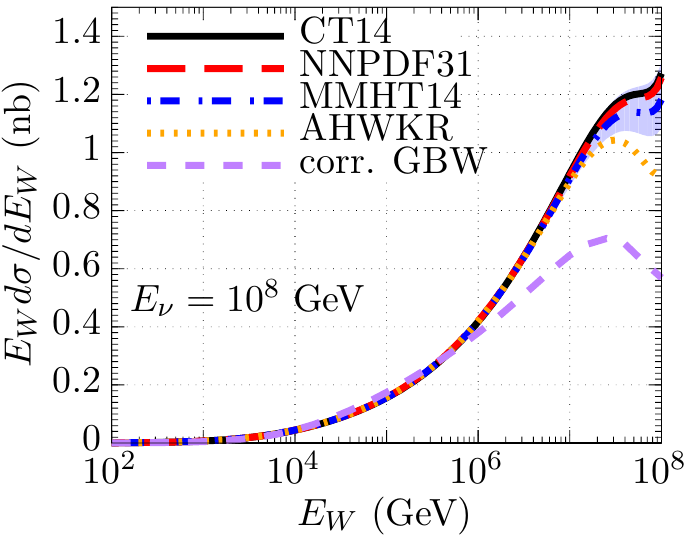}
		\caption{\sf \small {DIS neutrino-nucleon (isoscalar) differential cross-section in the $W$-boson energy for NNLO PDFs CT14~\cite{Dulat:2015mca}, MMHT14~\cite{Harland-Lang:2014zoa} (with uncertainty band), and NNPDF3.1~\cite{Ball:2017nwa} and color dipole models AHWKR~\cite{Arguelles:2015wba} and corrected GBW~\cite{GBW:1998} for a neutrino laboratory-frame energy of $E_\nu = 10^{8}\,$GeV. }}
		\label{fig:nuPdiff8}
	\end{center}
\end{figure}

\begin{figure} [htb]
	\begin{center}
		\includegraphics[clip=true,trim=0.0cm .0cm 0.0cm 0.0cm,width=12.0cm]{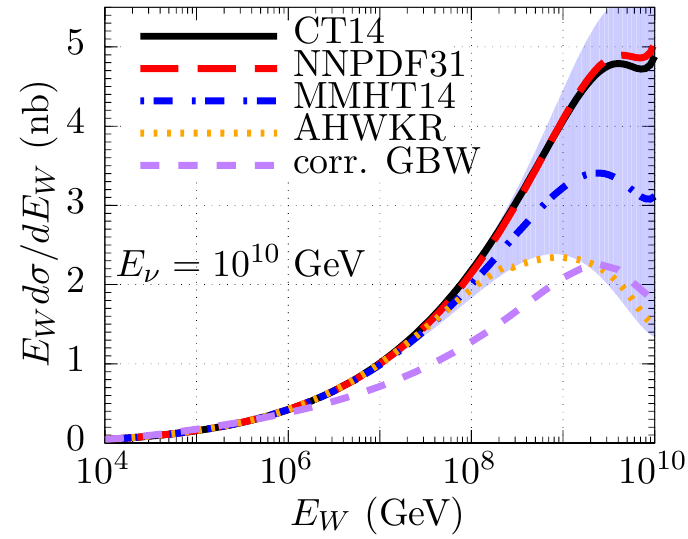}
		\caption{\sf DIS neutrino-nucleon (isoscalar) differential cross-section in the $W$-boson energy for NNLO PDFs and color dipole models for a neutrino laboratory-frame energy of $E_\nu = 10^{10}\,$GeV.}
		\label{fig:nuPdiff10}
	\end{center}
\end{figure}

\begin{figure} [htb]
	\begin{center}
		\includegraphics[clip=true,trim=0.0cm .0cm 0.0cm 0.0cm,width=12.0cm]{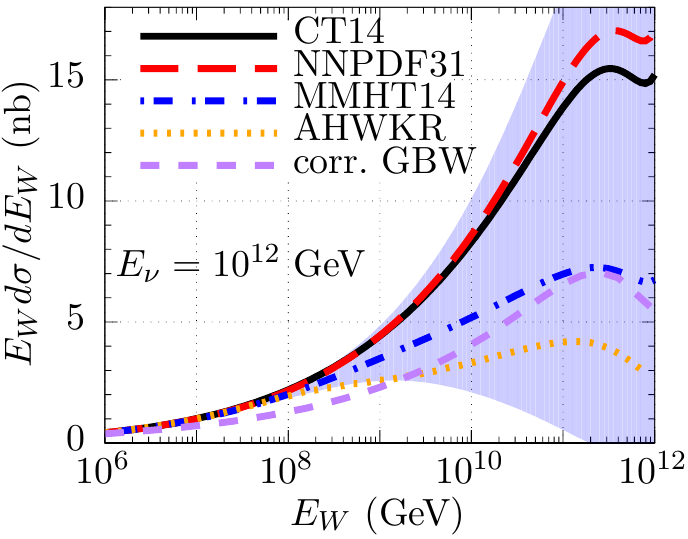}
		\caption{\sf Differential cross-section of DIS neutrino-nucleon in the $W$-boson energy for NNLO PDFs and color dipole models for a neutrino laboratory-frame energy of $E_\nu = 10^{12}\,$GeV.}
		\label{fig:nuPdiff12}
	\end{center}
\end{figure}

In Figs.~\ref{fig:nuPdiff8}, \ref{fig:nuPdiff10}, and \ref{fig:nuPdiff12} we show the differential cross section obtained from Eq.~\ref{eq:diffXY} once integration in $x$ is performed in the appropriate kinematical range. Again we analyze the three PDF sets aforementioned (CT14, MMHT14, and NNPDF3.1) and the two dipole models (GBW and AHWKR) at three neutrino energies: $E_\nu = 10^8$, $10^{10}$, and $10^{12}$\,GeV. The figures are built in such a way that the contribution to the total cross section (area under the curve) can be easily seen ($E_W d \sigma / d E_W$ by log scale in $E_W$). 

In Fig.~\ref{fig:nuPdiff8}, we see practically no difference among the parton distribution functions and dipole models for neutrino energy up to 1 PeV. However, dipole models show some difference, for instance, when the first bang energy is $E_{W} \gtrsim 10^{7}$ GeV. This behaviour can be easily understood by noting that, at this energy, $x_\text{min} \sim 10^{-6}$ and therefore the dynamics of DIS starts to be dominated by very low-x partons. In this respect, it is also worth noticing the smooth transition of the AHWKR model from pure DGLAP evolution to the saturation regime, starting at $\sim 3 \times10^{7}$ GeV. These differential cross sections show that information from a measurement of double bang events will shed some light into the proton structure, as we found significant discrepancies among the extrapolations (specially for large inelasticities $ 0.3 \lesssim y < 1$). When integrated, small and large $E_{W}$ behaviours lead to a total $\nu N$ cross section from different models that are very similar. Therefore, energy distributions such as the one displayed yield much more clear signatures of parton saturation then integrated cross sections already at a lower neutrino energy since for large inelasticities predictions of these models are not compatible with MMHT14 uncertainty band, as Fig.~\ref{fig:nuPdiff8} shows.

Moving towards the very high energy frontier, one analyses also $d\sigma/dE_{W}$ at higher neutrino energies, like $E_{\nu} = 10^{10}$ and $10^{12}$\,GeV. Our results are displayed in Figs.~\ref{fig:nuPdiff10} and \ref{fig:nuPdiff12}. In comparison to Fig.~\ref{fig:nuPdiff8}, these results are even more sensitive to different low-$x$ extrapolations, and uncertainties in MMHT14 predictions are as well enhanced. CT14 and NNPDF3.1 produce very similar results (within a few percent) in both cases, while MMHT14 central result is less than half of the others at $10^{12}$ GeV; and is actually closer to the saturation models GBW and AHWKR. A very interesting aspect of this results is that it clearly shows that DGLAP partons can show a behaviour very close to the one showed by dipole models. 

Dipole models produce even smaller cross sections at large $E_W$ with increasing neutrino energy.  In effect, we can see from Figs.~\ref{fig:nuPdiff10} and \ref{fig:nuPdiff12} that, not only the threshold for parton saturation moves towards higher $E_{W}$ energies, but also their magnitudes relative to the corresponding cross section values predicted by PDFs. So, in case of clear identification of double bang events, the energy evolution of the magnitude of the signal observed might be used to discriminate among various scenarios of low-x QCD evolution and to narrow down the uncertainties of PDFs in the low-x region. Despite the uncertainities make all five curves compatible with MMHT14 close to $E_{W}=10^{12}\,$ GeV, the dipoles and PDF models are incompatible under $1\sigma$ in the $10^{8}-10^{9}$ GeV in the E$_{W}$ range, which are values more attainable to be measured in the next generation of neutrino detectors.

In Figs.~\ref{fig:nuPdiff8}, \ref{fig:nuPdiff10}, and \ref{fig:nuPdiff12} it is possible to identify a change of shape of the PDFs lines for $E_{W}\rightarrow E_{\nu}$. Several specific details of the parton distributions (specially the sea quarks but also including valence quarks) dominate the cross section behaviour in this region. Then, there is no single explanation for the observed effect, as its due to the interplay of high $Q^{2}$ and small $x$ in the evolution of PDFs in the threshold $E_{W}\approx E_{\nu}$.

\begin{figure} [htb]
	\begin{center}
		\includegraphics[clip=true,trim=0.0cm .0cm 0.0cm 0.0cm,width=12.0cm]{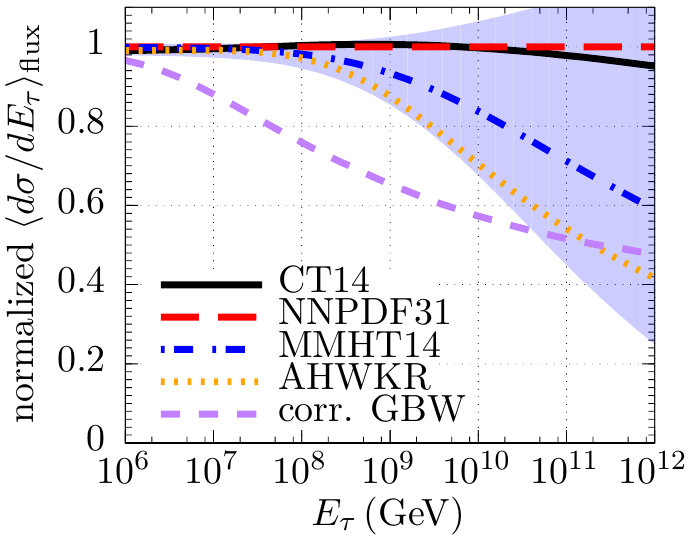}
		\caption{\sf Differential cross section for suggardaddy and lollipop events as a function of tauon energy. All curves have been normalized by the NNPDF3.1 result. Uncertainties ($1\sigma$) in the MMHT14 prediction are given by the blue band.}
		\label{fig:sugar}
	\end{center}
\end{figure}

Still considering an incoming tau neutrino, if the first bang is not detected but the second one is,  we have a suggardaddy or a lollipop event. Therefore, one can be sure about the tauon energy, but there will be little or no information about the neutrino energy. Now the right choice is to integrate over the neutrino flux and the price to pay is that now we have some uncertainty due to lack of precise knowledge about that flux. In this context, a possible observable to look at would be another differential cross section, in this case as function of the $\tau$ energy, namely:
\begin{align}
\label{eq:diffEtau}
\left\langle \frac{d \sigma}{d E_\tau } \right\rangle_\text{flux}
= \int_{E_\tau}^\infty d E_\nu   \frac{d f_\nu}{d E_\nu}
\int_{x_\text{min}}^1 d x \, \frac{1}{E_\nu} \frac{d \sigma}{d y d x}.
\end{align}
As before, $x$ integration is performed with the differential cross section given in Eq.~\ref{eq:diffXY}, resulting in a differential spectrum for $y$; that in turn shall be integrated over the average flux considering  $E_\nu > E_\tau$.  

In Fig.~\ref{fig:sugar}, we show our calculations of the differential cross section in Eq.~\ref{eq:diffEtau}, using the same PDF sets and dipole models used before. To have clear and direct comparison between them, we normalize, at each tauon energy, all results  by the NNPDF one. For tauon energies larger than $10^9$ GeV, all calculated curves produce very distinct results, meaning that depending on the number of gluons predicted by a particular DGLAP evolved set or a particular dipole model, we will see more or less lollipop and sugardaddy events. An equivalent way of understanding these results is to look at the slope of these curves. Models in which the proton structure (gluon number or dipole cross section) grows faster at decreasing $x$ or at increasing $Q^2$ will have a more positive (less negative) slope. Put in other words, for models which include parton saturation, the relative probability (density) of finding a second bang inside the detector region becomes significant smaller. For instance, for very energetic tauons, typically with $E_{\tau}=10^{9}$ GeV, AHWKR and GBW predictions can be some $15-35\%$ lower than pure DGLAP dynamics provide. 

An important feature of these results is small sensitivity to changes in the neutrino flux. In fact, we have tested some variations in the neutrino-energy dependence of the neutrino flux by allowing the exponent $\delta$ in Eq.~\ref{eq:flux} to vary in the range $0.2-0.4$ and found that our curves do not change much in this respect. This is mainly due to the fact the integration in Eq.~\ref{eq:diffEtau} is dominated by the lower energy neutrinos, as the neutrino flux is reasonable hard. However, a problem may arise at energies of $E_{\tau}>10^{9}$, as in this case the tauon will travel on average more than the atmosphere thickness, and the measurement of such kind of events will only happen at small angles above the horizon, reducing the total number of them.

\begin{figure} [htb]
	\begin{center}
		\includegraphics[clip=true,trim=0.0cm .0cm 0.0cm 0.0cm,width=12.0cm]{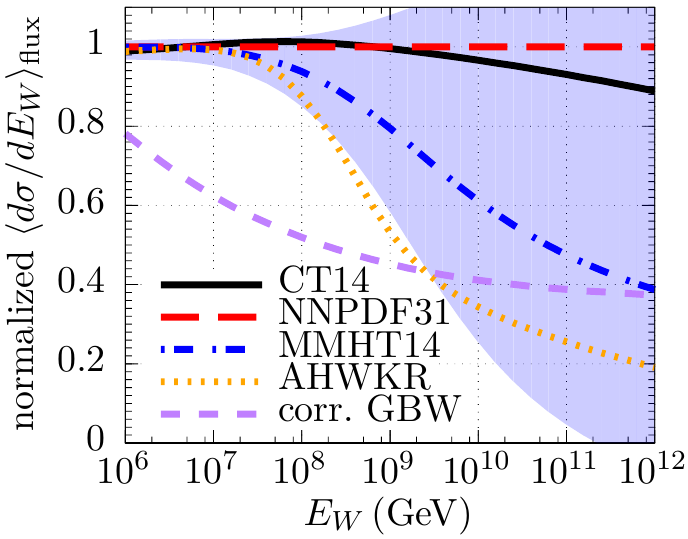}
		\caption{\sf Differential cross section for inverted lollipop events as a function of W boson energy. All curves have been normalized by the NNPDF3.1 result. Uncertainties ($1\sigma$) in the MMHT14 prediction are given by the blue band. }
		\label{fig:inverted}
	\end{center}
\end{figure}

Next, following previous discussions, we propose a possible analysis of inverted lollipop events, in which the first bang happens inside the detector and the second bang is not detected. Again it is necessary to integrate over the neutrino energy considering the incoming flux, but this time it is the W boson energy that is fixed. 

In Fig.~\ref{fig:inverted}, we show our results for the normalized cross section as a function of W boson energy. Compared to Fig.~\ref{fig:sugar}, we see more severe discrepancies among the central predictions of various models. Interestingly, deviations from pure DGLAP behaviour start a little earlier, at $10^7$ GeV (one order of magnitude earlier  than the other topologies). This is expected though, as the steep decrease of the flux will cut off large $E_{\nu}=E_W + E_\tau$ contributions. Therefore, the curves one shows in Fig.~\ref{fig:inverted} are dominated by small $E_{\nu}$ energy; and with $E_W$ fixed, $E_\tau$ shall be small compared to $E_W$ in order to probe the small $x$ region. As before, the detection of this class of event would be very informative for studies of parton saturation, as it could reveal, already at a few PeV, severe discrepancies among scenarios with or without gluon saturation. 

Concerning the uncertainty bands of MMHT14, we see that dipole predictions can fall outside them. This is a sufficiently large discrepancy that will require one of the models (dipoles or partons) to be at least adapted, if not rethought. Even if we think only about the PDFs, the uncertainty bands show that measurements at the relatively low energy of $E_W = 10^6$ -- $10^7$\,GeV in the case of an inverted lollipop event can probe PDFs where the cross section theoretical error is of the order of 5\%, much higher than the 1\% precision of typical current collider calculations.

\section{Conclusion}

In this work we have studied the interaction of astrophysical neutrinos with matter. We started by calculating the neutrino-nucleon charged current total cross section, $\sigma_{\nu N}$, in a large energy range (spanning from a few PeV up to EeV) using  two classes of models: partons and color dipoles. As expected, screening effects due to parton saturation at small $x$ play a major role in the cross section, already at $10^9\,$GeV. However, in spite of being an important observable to distinguish various models of QCD interactions and reduce uncertainties in the parton distributions, at the energies in which there are more discrepancy among models and more uncertainties is also where the flux of astrophysical neutrinos is expected to be very small.   

Having identified this problem, we recognize the analysis of differential cross sections in W-boson energy as a viable observable in the special case of double bang events. In this context, we found large differences between PDF and dipole models, specially in the region of mid to large inelasticities, $0.3\lesssim y < 1$, already at lower neutrino energies such as $E_{\nu}=10^{8}$ GeV. We also show plots for energies of $10^{10}$ and $10^{12}$ GeV, as our main goal in this study is to demonstrate the potential usefulness of these results in discriminating among some scenarios of parton saturation and usual DGLAP dynamics. Our findings qualify the differential observable  $d\sigma/dE_{W}$ as a proper tool to investigate low-x QCD processes, due to the fact that double bangs can be measured in Cherenkov radiation based detectors in the near future. 

Nonetheless, we propose a parallel investigation of events in which only one bang is detected, such as lollipops, sugardaddies and inverted lollipops, by following a similar approach. In order to do that, we integrate the differential cross section over the neutrino energy, including the neutrino flux but keeping the dependence in $E_\tau$ (as it can be measured in events with lollipop and sugardaddy topologies). The same can be done for inverted lollipop events, although in this case the variable to be kept fixed is the first bang energy ($E_W$). Interestingly, we found significant discrepancies among the central predictions of the various NNLO PDFs tested (NNPDF3.1, CT14 and MMHT14) and dipole models in all results, even at a relatively low neutrino energy of $10^6$\,GeV. This demonstrates that such approach can indeed be used to better understand QCD dynamics at very low-x and reduce PDFs uncertainty bands.  

In conclusion, the measurement and detailed study of high energy tau neutrino interactions like double bang, lollipop, sugardaddy and inverted lollipop events can provide new information on the proton (nucleon) structure at small $x$ in a not so distant future. More important, it can be achieved in astrophysical neutrino experiments, at energies not reachable by current colliders. This knowledge will be, of course, of great value not only in revealing essential features of QCD dynamics at the PeV-EeV scale and beyond, but also in studying potential backgrounds for new physics.

\section*{Acknowledgements}

We are grateful to Bruna de Oliveira Stahlh\"ofer and Roman Pasechnik for the critical reading of this manuscript and valuable discussions. This work was supported by CNPq, Capes, Fapesc for EGdO. This study was financed in part by the Coordena\c{c}\~ao de Aperfei\c{c}oamento de Pessoal de N\'ivel Superior -- Brasil (CAPES) -- Finance Code 001. EGdO and DAF also acknowledges support by the project INCT-FNA (464898/2014-5).

\end{document}